\documentclass[sigconf]{acmart}
\AtBeginDocument{%
  }

\setcopyright{none}  
\renewcommand\footnotetextcopyrightpermission[1]{}  

\acmConference[Augmented Educators and AI(CHI 2025 Workshop)]{Augmented Educators and AI: Shaping the Future of Human-AI Collaboration in Learning on CHI 2025 Workshop}{April 26,2025}{Yokohama, JAPAN}

\usepackage{array}

\begin{document}

\title{Design of AI-Powered Tool for Self-Regulation Support in Programming Education}

\author{Huiyong Li}
\orcid{0000-0001-9916-7908}
\affiliation{%
  \institution{Research Institute for Information Technology, 
  Kyushu University}
  \city{Fukuoka}
  \country{Japan}
}
\email{li.huiyong.194@m.kyushu-u.ac.jp}

\author{Boxuan Ma}
\authornote{Corresponding author.}
\orcid{0000-0002-1566-880X}
\affiliation{%
  \institution{Faculty of Arts and Science, Kyushu University}
  \city{Fukuoka}
  \country{Japan}}
\email{boxuan@artsci.kyushu-u.ac.jp }

\renewcommand{\shortauthors}{Li \& Ma}

\begin{abstract}

Large Language Model (LLM) tools have demonstrated their potential to deliver high-quality assistance by providing instant, personalized feedback that is crucial for effective programming education. However, many of these tools operate independently from institutional learning management systems, which creates a significant disconnect. This isolation limits the ability to leverage learning material and exercise contexts for generating tailored, context-aware feedback. Furthermore, previous research on LLM support for programming learning mainly focused on knowledge acquisition, not the development of important self-regulation skills. To address these challenges, we designed CodeRunner Agent, an LLM-based programming tool that integrates the CodeRunner, a student-submitted code executing and automated grading plugin in Moodle. CodeRunner Agent enhances students' contextual self-regulated learning by providing learning log-based contextual feedback and self-regulation strategy-based AI scaffolding. Additionally, CodeRunner Agent empowers educators to customize AI-generated feedback by incorporating detailed context from lecture materials, programming questions, student answers, and execution results. This integrated, context-aware, and skill-focused approach offers a promising avenue for data-driven programming education.

\end{abstract}





\maketitle
{\small
\noindent ©  This paper was adapted for the \textit{CHI 2025 Workshop on Augmented Educators and AI: Shaping the Future of Human and AI Cooperation in Learning},
held in Yokohama, Japan on April 26, 2025. This work is licensed under the Creative Commons Attribution 4.0 International License (CC BY 4.0).
}

\section{Introduction}

Programming has become an increasingly important part of university education; however, it is becoming more challenging for educators to provide timely, personalized support to each student\cite{chiu2024impact,yilmaz2023effect,ma2024enhancing}. Traditional support methods, such as scheduled office hours, are often limited, and in-person help can be both time-consuming and labor-intensive. In this context, Large Language Models (LLMs) are emerging as a promising solution to this challenge, offering the potential for on-demand, personalized programming support that can supplement traditional teaching methods\cite{yilmaz2023augmented,pankiewicz2023large}. A variety of LLM-powered programming assistants have been developed to provide timely coding guidance and suggestions, potentially transforming the landscape of programming education. 

While these advancements present exciting opportunities for personalized learning and efficient problem-solving, they also raise critical questions about the role and effectiveness of LLM-powered assistants in truly enhancing student learning. One major concern is that students may become overly reliant on these tools, potentially hindering the development of self-regulated learning (SRL) skills and problem-solving skills \cite{gong2024impact,rajala2023call,humble2023cheaters}.
Research suggests that the convenience of receiving direct answers from LLMs may hinder the development of SRL skills, as students may avoid the deeper cognitive effort required to work through challenges independently\cite{prasad2024self,skjuve2023user,tlili2023if}. This harmful effect becomes more serious in the introductory programming in university since it’s a challenging process in programming learning for freshman students. Unfortunately, most existing LLM-powered programming tools fail to address this issue and ignore the implementation of pedagogically sound scaffolding to enhance students' self-regulated learning while avoiding LLMs' harmful effects\cite{becker2023programming,sun2024would}. 

Another significant issue is that many LLM-based tools operate independently of institutional Learning Management Systems (LMS) like Moodle. This separation creates a disconnect between the tool and the broader educational context such as course materials and rich assignment details. For example, Ma et al. \cite{ma2024enhancing} found that when LLMs are not aligned with a student's specific curriculum, they generate advanced or off-topic responses that stray from the intended teaching scope, ultimately detracting from the learning process. Without seamless integration, educators struggle to track how students interact with LLM-generated feedback over time, making it difficult to assess the true impact on learning outcomes \cite{ma2024exploring}. Therefore, it's essential to ensure that the AI-generated feedback is both relevant and aligned with course objectives by incorporating LLM-based tools within the LMS environment. The lecture materials and exercises in programming education are highly important contextual information; however, they are not well connected in current LLM-based scaffolding research \cite{kazemitabaar2024codeaid}. The Learning Analytics (LA) technique can enhance the quality of LLM-based feedback by utilizing the behavioral data in context between learners, lecture materials, and exercise solving \cite{kaur2024learning}. 

To address these limitations, we designed CodeRunner Agent, a LLM-based programming tool that seamlessly extends CodeRunner, a free and open-source plug-in in Moodle designed to execute and assess student-submitted code. CodeRunner Agent is designed to meet the needs of both learners and educators. It leverages the comprehensive context available within an LMS environment from learning logs and enhances students' self-regulated learning in introductory programming education. Specifically, it enhances the delivery of individualized feedback by combining students' knowledge level, self-regulated behaviors, and strategy-focused LLM-based scaffoldings. Beyond contextual and strategy-based feedback, CodeRunner Agent tracks students' requests and AI responses. This capability provides educators and researchers with valuable insights into AI-powered learning process and the overall effectiveness of AI-assisted instruction. By collecting and analyzing data on how students interact with AI-generated feedback, our approach paves the way for data-driven improvements in programming education. In summary, our proposal represents a significant step forward in integrating AI tools within institutional LMS environments to enhance programming education. By addressing the challenges of AI integration, contextual feedback, and self-regulation scaffolding, our work offers promising avenues for enhancing student skill development and deepening our understanding of AI’s role in modern education.

\section{Related Work}
 
As LLMs become increasingly pervasive, educational researchers are examining their potential to generate educational content, boost student engagement, and personalize learning experiences. This is particularly relevant in programming education, where the adoption of such tools is prompting a reevaluation of traditional teaching methods \cite{shoufan2023exploring,kasneci2023chatgpt}. 

Recent studies have primarily focused on assessing LLMs' capabilities in programming tasks, ranging from code generation and program repair to code explanation and code summarization \cite{rajala2023call,chen2023gptutor}. For instance, Finnie-Ansley et al. \cite{finnie2023my} demonstrated that OpenAI Codex outperforms most students on code-writing questions in both CS1 and CS2 exams. In a similar vein, Savelka et al. \cite{savelka2023thrilled} evaluated GPT-3 and GPT-4 on programming exercises across three Python courses, revealing that these models progressed from failing typical assessments to passing courses without human intervention. Furthermore, Sarsa et al. \cite{sarsa2022automatic} examined programming exercises generated by OpenAI Codex, assessing their novelty, plausibility, and readiness, and highlighted the potential for these models to create effective coding assignments. Recent work by Phung et al. \cite{phung2023generative} systematically compared GPT models with human tutors, finding that they approach human-level performance in both Python programming tasks and the resolution of real-world buggy programs. Additionally, ChatGPT has proven effective in providing feedback on programming assignments and aiding students in applying theoretical knowledge practically. Prior research has underscored the model's capacity to generate personalized feedback that students rate positively \cite{pankiewicz2023large}.

While LLMs hold significant promise for enhancing programming education, both students and researchers have expressed concerns about their direct use. One major worry is that students might become overly reliant on LLMs, potentially stunting the development of SRL skills. Additionally, many students struggle with formulating effective prompts, often resulting in feedback that fails to meet their learning needs. In response, researchers have increasingly developed specialized LLM-based tools that address these issues. For instance, Kazemitabaar et al. \cite{kazemitabaar2023studying} developed Coding Steps, which leverages LLM-based code generators to support beginners in introductory programming courses. Similarly, Lifton et al. \cite{liffiton2023codehelp} introduced the CodeHelp tool, designed to assist students while incorporating guardrails that prevent the tool from directly revealing complete solutions. With CodeHelp, students can input a free-form question along with their code and, optionally, an error message, ensuring that the feedback remains contextual and instructive. In another example, CodeAid \cite{kazemitabaar2024codeaid} offers a range of input templates and interactive response formats tailored to diverse student needs. It employs scaffolding techniques, such as interactive pseudo-code and detailed code annotations, to guide students from grasping fundamental programming concepts to independently writing and debugging their code. These innovations collectively illustrate the emerging trend of developing LLM-based tools that not only harness the power of AI but also promote deeper learning and independence among students.

Despite the impressive achievements of these tools in enhancing programming education, they often operate in isolation, lacking the integration of various functionalities and failing to connect with existed Learning Management Systems (LMS). This fragmented approach does not adequately meet the needs of educators and students, thereby hindering widespread adoption and scalability. To address these limitations, we developed CodeRunner Agent, a comprehensive solution that seamlessly integrates LLM-powered assistance with LMS platforms. By unifying these functionalities, our system offers a more cohesive and effective learning environment, ensuring that both instructors and learners have access to timely, context-aware support on a large scale.

\section{CodeRunner Agent}

\subsection{Overview of CodeRunner Agent}

\begin{figure*}[ht]
\centering
\includegraphics[scale=0.55]{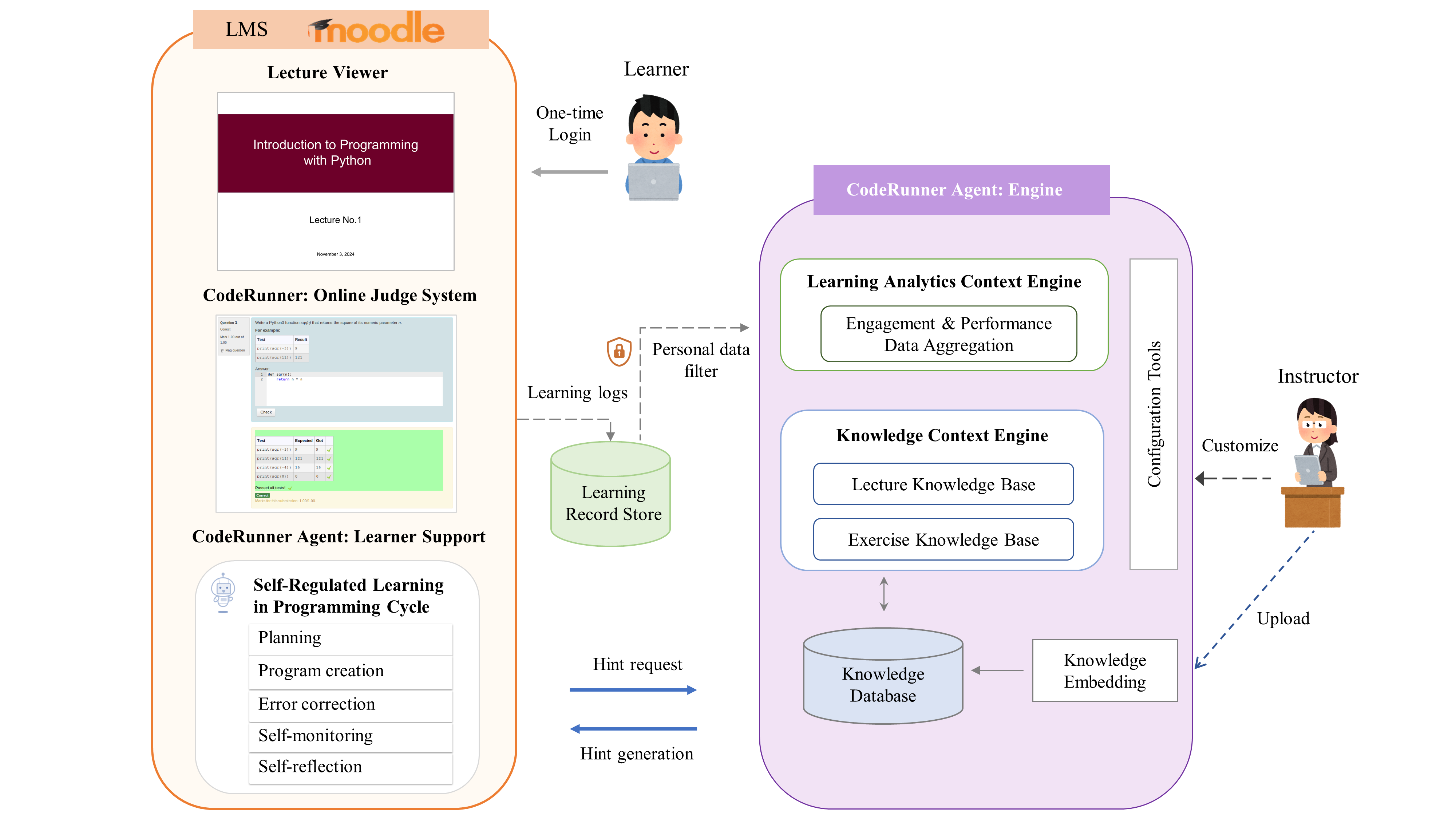}
\caption{Overview of Programming Support Environment}
\label{fig:1}
\end{figure*}

The framework of CodeRunner Agent is shown in Figure \ref{fig:1}. It is designed and developed to scaffold programming education with an LLM-based tool in the Moodle LMS. The framework contains a lecture viewer, a CodeRunner plugin and CodeRunner Agent.

The lecture viewer is provided to deliver the lecture slides by instructors and access the lecture slides by learners inside and outside of class. The operations of the lecture viewer are recorded in the form of Experience API (or xAPI) statements. Then the xAPI statements are stored in the Learning Record Store (LRS). The examples of the operation logs are accessing time and accessing frequency.

CodeRunner \cite{lobb2016coderunner} is a free, open-source plugin and it can be imported to Moodle LMS. Learners can write programming codes to solve programming problems and receive automated grades by running it in a series of tests. A logging plugin named “Logstore xAPI” \cite{rotelli2023moodle} is used to record the CodeRunner results and send them to the LRS. For example, the code for the test, got output, correct status, mark reward will be logged in LRS if learners run the test code once for testing in CodeRunner.

CodeRunner Agent is an AI-powered tool to support learners' SRL and teachers' customized designs for LLM-based feedback in programming education. It can be embedded into the Moodle LMS and executed with the CodeRunner plugin. Learners can receive general-purpose and programming-specific regulatory strategy hints from LLMs-based feedback. Instructors can upload learning materials to the context engine of the CodeRunner Agent and customize the contextual parameters of the agent to create tailored instruction. The interactions with the CodeRunner Agent are automatically tracked as xAPI statements and stored in the LRS. For instance, the request type, request time, and exercise ID related to the request will be recorded in LRS if learners send a request to the agent.

\subsection{SRL Support Model in CodeRunner Agent}

The SRL scaffoldings in CodeRunner Agent are implemented using a five-phase cycle model named PPESS: \textit{Planning, Program creation, Error correction, Self-monitoring, Self-reflection}. The PPESS model is grounded in the well-known Zimmerman's SRL theory \cite{zimmerman2008investigating} and specified for self-regulation in programming learning \cite{silva2024learning}. 

The phase, target SRL strategy, and LLMs support in the CodeRunner Agent are summarized in Table \ref{table:1}. Both general-purpose and programming-specific regulatory strategies are included for SRL scaffoldings in the five-phase PPESS model. 

The \textit{Planning} phase serves as the foundational stage in programming, including cognitive and metacognitive strategies for problem understanding, problem decomposition, solution architecture and essential programming component identification. The \textit{Program creation} phase represents an implementation stage, wherein learners execute preliminary plans by implementing programming constructs that address identified requirements. This phase involves the combination of declarative and procedural knowledge within the programming task, facilitating the transfer of coding knowledge. The \textit{Error correction} phase encompasses sophisticated  learning strategies for addressing coding errors. These include the externalization of metacognitive processes to remediate conceptual misunderstandings, employment of cognitive diagnosis strategies for systematic problem identification, and implementation of self-regulation to optimize the debugging process. The \textit{Self-monitoring} phase occurs concurrently with the program creation and error correction phases, representing an ongoing metacognitive process. Learners engage in continuously assessment and adjustment toward task completion and code quality. Finally, the \textit{Self-reflection} phase, occurring at the end of programming, enables learners to comprehensively evaluate their programming experience by identifying areas requiring improvement, maintaining motivation, and improving current learning strategies for future tasks.

\begin{table*}[htb]
  \caption{Phase, Target SRL strategy, LLM-powered feedback in the CodeRunner Agent.}
  \label{table:1}
  \begin{tabular}{  p{8em}  p{6cm} p{8cm}  }
    \toprule
    Phase&Strategy&LLM-powered feedback\\
    \midrule
    Planning & Problem understanding, problem definition, program logic planning  & Offer the basic knowledge of the exercise’s requirements and suggest planning the program step-by-step using diagrams, pseudocode, or notes. \\
    Program creation & Review lecture materials, review previous exercises, code dividing, code commenting & Provide the location of required knowledge in lecture materials, supplemental resources related to the exercise, and explanations for the key points. \\
    Error correction & Review the exercise statement, utilize test cases, analyze the error message, help-seeking & Give suggestions for effective error correction and generate hints on fixing syntactic and logical errors without showing the solution directly. \\
    Self-monitoring & Check exercises progress, test the program regularly & Encourage learners to track their own learning progress regularly. \\
    Self-reflection & Achievement self-assessment, effort self-assessment, code review, code optimization  & Provide evaluations on learners' behavioral process and final performance,  motivate learners by finding their strength points or the effort they put in, and suggest learners to identify their improvement areas.  \\
  \bottomrule
\end{tabular}
\end{table*}

\begin{figure}[t]
\centering
\includegraphics[scale=0.6]{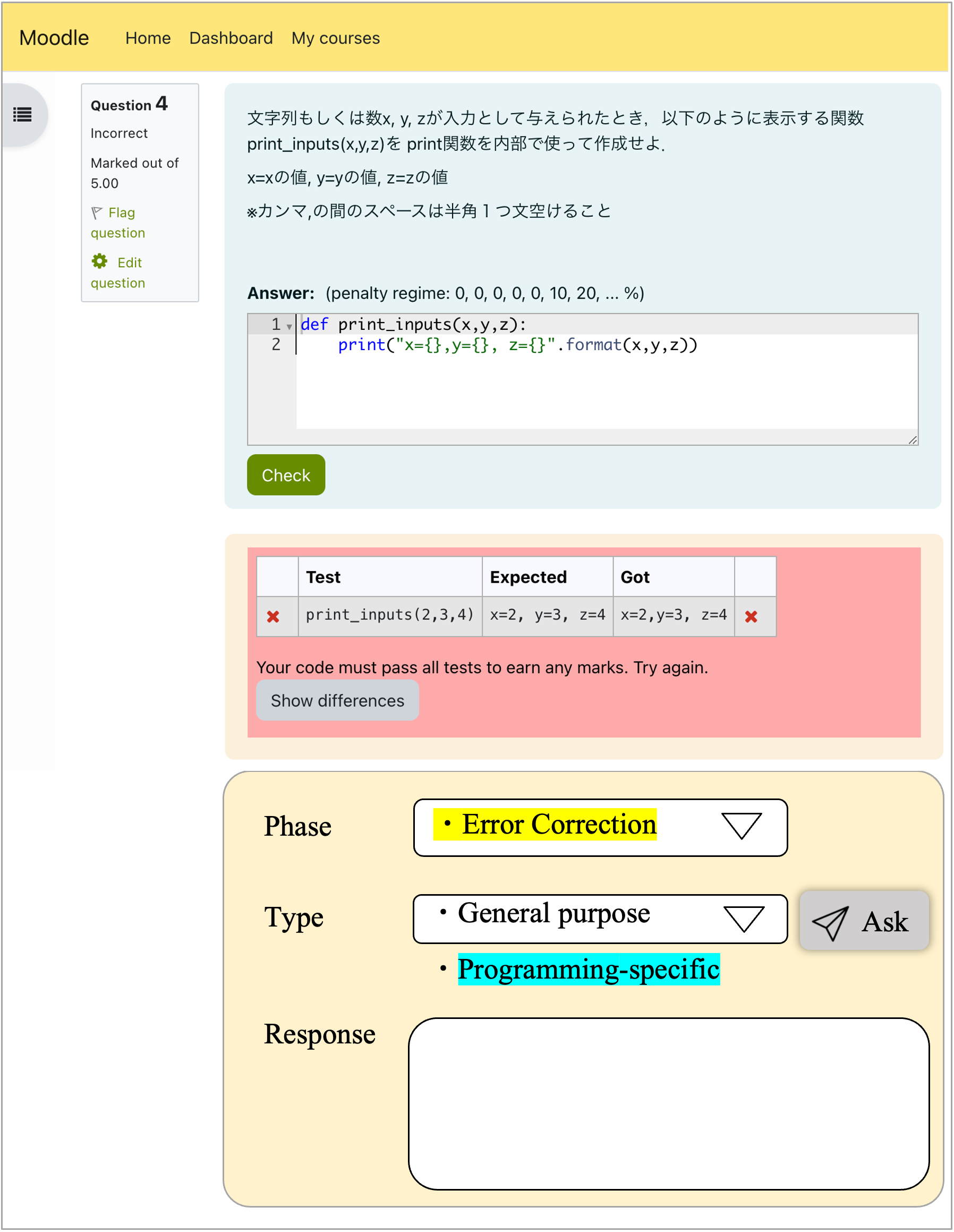}
\caption{User Interface of Integrated CodeRunner Agent for Learners}
\label{fig:2}
\end{figure}

The user interface of the integrated CodeRunner Agent for learners is shown in Figure \ref{fig:2}. It contains three main components: (a) Question \& Answer, (b) Check with Test Cases, and (c) LLM-based Support. Firstly, learners check the question statement for programming exercise and input their code solutions for the question in the top Question \& Answer component. Secondly, learners submit their code solutions for checking pre-defined test cases by comparing the execution results with the expected outputs in the central Check with Test Cases component. Finally, learners have the option to select one of five SRL phases like error correction, choose one request type from general purpose or programming-specific, and send the request to the agent in the bottom LLM-based Support. Upon submitting a request, the LLM-based strategy-focused feedback is displayed within the response box.

\subsection{Context Engine in CodeRunner Agent}

The CodeRunner Agent's core intelligence contains two context engines: Learning Analytics Context Engine (LACE) and Knowledge Context Engine (KCE). LACE calculates and aggregates learners' engagement metrics (i.e., time spent, attempt frequency) and performance metrics (i.e., success rates, error patterns) from LRS data after filtering personal data (i.e., name, gender, and email). LACE is integrated as contextual embedding of self-regulation behavioral strategies with the feedback of the agent. KCE handles both lecture and exercise knowledge bases. The lecture knowledge base manages the key programming concepts and the dependencies between the concepts from lecture materials. The exercise knowledge base categorizes exercises by concept, difficulty, solution, and typical mistakes. KCE is used as learner knowledge embedding in the agent.
 
Instructors can upload lecture materials (i.e., concept definition, concept illustration, and annotated examples) and exercises (i.e., problem statement, solution, and test cases) to the context engine. They will be converted to textual knowledge and stored in the knowledge database for future retrieval. More importantly, instructors can update the knowledge bases and customize the parameters of the context engine by using the configuration tools.

\section{Discussion and Future work}

This study aims to design an LLM-based programming tool, CodeRunner Agent, that seamlessly integrates with a lecture viewer and CodeRunner plugin in the Moodle LMS. The design is grounded within Zimmerman's SRL theory and targets the freshmen students' programming-specific knowledge acquisition as well as general self-regulation skill development. The agent provides learning log-based contextual feedback and self-regulation strategy-based AI scaffolding to enhance contextual SRL. Furthermore, the agent empowers educators to customize AI-generated feedback by incorporating detailed context from lecture materials, programming questions, student answers, and execution results. This study can fill the gap between AI, SRL, and LA by combining learning log-based contextual feedback, LLM-based self-regulation scaffolding, and seamlessly integrating in LMS. 

Future work will be conducted for the interface improvement of the configuration tools and user-friendly workflow in the instructor-agent collaboration. Additionally, the effects of the CodeRunner Agent will be evaluated through short-term pilot studies in an actual class and semester-long experiments in multiple actual classes in university.

\bibliographystyle{ACM-Reference-Format}
\bibliography{base}

\end{document}